\documentstyle[12pt]{article}
\title{
Model Representation for Self - Consistent - Field Theory of
Isotropic Turbulence}
\author{R.~V.~R.~Pandya\\
Department of Mechanical Engineering \\University of Puerto Rico
at Mayaguez\\
 Mayaguez, Puerto Rico, PR 00681, USA}

\begin{document}

\maketitle

\begin{abstract}
In this paper, Langevin model equation is proposed for Fourier
modes of velocity field of isotropic turbulence whose statistical
properties are identical to those governed by equations of
Self-Consistent-Field (SCF) theory of turbulence [J. R. Herring,
Physics of Fluids {\bf 9}, 2106 (1966)].
\end{abstract}

\section{Introduction}

Kraichnan's seminal and pioneering work on Direct Interaction
Approximation (DIA) (\cite{Kraichnan58}, \cite{Kraichnan59}) and
Lagrangian History Direct Interaction Approximation (LHDIA)
(\cite{Kraichnan65a}) has been influential in setting the right
tone in the field of theory of turbulence and leading to other
fundamental renormalized approaches for turbulence closure
(\cite{Edwards64}, \cite{Herring65, Herring66}, \cite{McComb78},
\cite{Kaneda81}, \cite{Lvov91}). Various renormalized approaches
are critically reviewed by \cite{Leslie73}, \cite{ McComb90},
\cite{McComb95}, \cite{Lvov91}, and \cite{Lesieur97}. And the
Self-Consistent-Field (SCF) approach of \cite{Herring66} having
close relationship with DIA (\cite{HK72}) is the central focus in
this paper.

\cite{Herring65} developed the SCF approach for stationary
isotropic turbulence and subsequently generalized to the
non-stationary isotropic turbulence (\cite{Herring66}). Instead of
applying the perturbation technique to the Navier-Stokes
equations, following \cite{Edwards64} framework, Herring preferred
Livouville equation for probability distribution function of
Fourier modes of the velocity field. Then, a self-consistent-field
procedure was carried out around the zeroth order probability
distribution which is the product of exact single mode
distribution. This led \cite{Herring66} to derive equations
governing time evolutions of Green's function, single-time
velocity correlation and two-time velocity correlation. Yet
another method of \cite{BS70} yielded the SCF set of equations and
thus doubly justified these equations. The equations for Green's
function and single-time velocity correlation are identical in
form to the corresponding DIA's equations and generalized
fluctuation-dissipation relation represents the equation for
two-time velocity correlation in the SCF approach.

Despite being closer to Edwards's theory framework and closer to
DIA in terms of the final equations, well justified SCF approach
lacks a model representation. Whereas model representations are
known to exist for DIA (\cite{Kraichnan70a}) and extended
Edwards's theory (\cite{Kraichnan71}) associated with the
non-stationary turbulence. Also, model representations are
available for Kaneda's theory (\cite{Kaneda81}) and
\cite{McComb78} local energy transfer (LET) theory
(\cite{Pandya04}). The model representation, if exists, assures
the fact that statistical properties predicted by SCF are those of
a realizable velocity field and consequently establishes certain
consistency properties. The purpose of this paper is to suggest an
existence of Langevin model representation for SCF. Consequently,
to make sure that SCF does not lag behind other theories when
judged from the perspective of realizability and model
representation.

\section{SCF theory equations}
In this section, closed set of equations describing the
statistical properties of isotropic turbulence as obtained by SCF
approach of \cite{Herring66} are presented. I should be excused
for not using the original notations of Herring, rather using the
notations of \cite{McComb90} while presenting SCF equations. The
Fourier modes $u_i({\bf k},t)$ defined by
\begin{equation}
u_i({\bf x},t)=\int d^3{\bf k}u_i({\bf k},t)\exp(i{\bf k}\cdot
{\bf x}),
\end{equation}
of the velocity field $u_i({\bf x},t)$ of homogeneous, isotropic,
incompressible fluid turbulence in space-time (${\bf x}-t$) domain
satisfy the following Navier-Stokes equation written in Fourier
wavevector ($\bf k$) and time domain:
\begin{equation}
(\frac{\partial}{\partial t}+\nu k^2)u_i({\bf k},t)= M_{ijm}({\bf
k})\int d^3{\bf p}u_j({\bf p},t)u_m({\bf k}-{\bf p},t).
\label{eq1}
\end{equation}
Here $\nu$ is kinematic viscosity of fluid, inertial transfer
operator
\begin{equation}
M_{ijm}({\bf k})=(2i)^{-1}[k_jP_{im}({\bf k})+k_mP_{ij}({\bf k})],
\end{equation}
the projector $P_{ij}({\bf k})=\delta_{ij}-k_ik_jk^{-2}$, $k=|{\bf
k}|$, and $\delta_{ij}$ is the Kronecker delta. The subscripts
take the values 1, 2 or 3 alongwith the usual summation convention
over repeated subscript. The two-time velocity correlation
$Q_{in}({\bf k},{\bf k'};t,t')=\langle u_i({\bf k},t)u_n({\bf
k'},t') \rangle$, single-time velocity correlation $Q_{in}({\bf
k},{\bf k'};t,t)=\langle u_i({\bf k},t)u_n({\bf k'},t) \rangle$ of
the velocity field $u_i({\bf k},t)$ and the Green's function
$G_{in}({\bf k};t,t')$ can be simplified for isotropic turbulence,
and written as
\begin{equation}
Q_{in}({\bf k},{\bf k'};t,t')=P_{in}({\bf k})Q(k;t,t')\delta({\bf
k}+{\bf k'}),
\end{equation}
\begin{equation}
Q_{in}({\bf k},{\bf k'};t,t)=P_{in}({\bf k})Q(k;t,t)\delta({\bf
k}+{\bf k'}),
\end{equation}
and
\begin{equation}
G_{in}({\bf k};t,t')=P_{in}({\bf k})G(k;t,t'),
\end{equation}
where $\langle \, \rangle$ represents ensemble average and
$\delta$ represents Dirac delta function. The SCF equation for
$G(k;t,t')$ may be written as
\begin{equation}
\left(\frac{\partial}{\partial t}+\nu k^2\right)G(k;t,t')+\int
d^3{\bf p}L({\bf k},{\bf p})\int_{t'}^t ds \,G(p;t,s)Q(|{\bf k
}-{\bf p}|;t,s)G(k;s,t')=0 \,\forall \,t>t' \label{scf1}
\end{equation}
and $G(k;t',t')=1$. The SCF equation for $Q(k;t,t)$ may be written
as
\begin{eqnarray}
\left(\frac{\partial}{\partial t}+2\nu k^2\right)Q(k;t,t)+2\int
d^3{\bf p}L({\bf k},{\bf p})\int_{0}^t ds \,G(p;t,s)Q(|{\bf k
}-{\bf p}|;t,s)Q(k;t,s)\nonumber \\
=2\int d^3{\bf p}L({\bf k},{\bf p})\int_{0}^{t} ds
\,G(k;t,s)Q(|{\bf k }-{\bf p}|;t,s)Q(p;t,s) \label{scf2}
\end{eqnarray}
where
\begin{equation}
L({\bf k},{\bf
p})=\frac{[\mu(k^2+p^2)-kp(1+2\mu^2)](1-\mu^2)kp}{k^2+p^2-2kp\mu}
\end{equation}
and $\mu$ is the cosine of the angle between the vectors ${\bf k}$
and ${\bf p}$. These equations (\ref{scf1}) and (\ref{scf2}) have
form identical to the corresponding equations obtained by DIA
theory. In SCF approach, the equation for $Q(k;t,t')$ is
associated with generalized fluctuation-dissipation relation
\begin{equation}
Q(k;t,t')=G(k;t,t')Q(k;t',t'), \,\, \forall t \ge t'. \label{scf3}
\end{equation}
We write the equation for $Q(k;t,t')$, by using equations
(\ref{scf1}) and (\ref{scf3}), in the following form convenient
for further use:
\begin{eqnarray}
\left(\frac{\partial}{\partial t}+\nu k^2\right)Q(k;t,t')+\int
d^3{\bf p}L({\bf k},{\bf p})\int_{0}^t ds \,G(p;t,s)Q(|{\bf k
}-{\bf p}|;t,s)Q(k;s,t')\nonumber \\
=\int d^3{\bf p}L({\bf k},{\bf p})\int_{0}^{t'} ds
\,G(p;t,s)Q(|{\bf k }-{\bf p}|;t,s)Q(k;s,t'). \label{scf4}
\end{eqnarray}
Thus equations ({\ref{scf1}), ({\ref{scf2}) and ({\ref{scf4}) form
a closed set of final equations of SCF approach of
\cite{Herring66}. Now the goal is to obtain model equation for
$u_i({\bf k},t)$ which would have statistical properties identical
to those as predicted by this closed set of equations. And a
Langevin equation as a model representation for SCF is presented
in the section to follow.

\section{Langevin model equation for SCF}

Similar to Langevin model representation for DIA, consider a
Langevin equation for $u_i({\bf k},t)$ written as
\begin{equation}
\left(\frac{\partial}{\partial t}+\nu k^2\right)u_i({\bf
k},t)+\int_{0}^t ds \,\eta (k;t,s)u_i({\bf k},s)=f_i({\bf
k},t)+b_i({\bf k},t) \label{lan1}
\end{equation}
where $\eta (k;t,s)$ is statistically sharp damping function,
$f_i({\bf k},t)$ is a forcing term with zero mean and $b_i({\bf
k},t)$ is white noise forcing term having zero mean. It should be
noted that $b_i({\bf k},t)$ is an additional new forcing term that
is not present in DIA's Langevin model representation. We consider
these two different forcing terms to be statistically independent
\begin{equation}
\langle f_i({\bf k},t)b_n({\bf k}',t') \rangle=0 \,\, \forall \,
{\bf k}', t' \label{prop1}
\end{equation}
and their statistical properties for isotropic turbulence,written
as
\begin{equation}
\langle f_i({\bf k},t)f_n({\bf k}',t') \rangle=P_{in}({\bf
k})F(k,t,t')\delta({\bf k}+{\bf k}') \label{prop2}
\end{equation}
and
\begin{equation}
\langle b_i({\bf k},t)b_n({\bf k}',t') \rangle=P_{in}({\bf
k})B(k,t)\delta({\bf k}+{\bf k}')\delta(t-t'). \label{prop3}
\end{equation}
For particular choice of $\eta (k;t,s)$, $F(k,t,t')$ and $B(k,t)$,
the Langevin equation (\ref{lan1}) would recover the closed set of
SCF equations (\ref{scf1}), (\ref{scf2}) and (\ref{scf4}). Now we
obtain that particular choice.

For isotropic turbulence, the Green's function of the Langevin
equation (\ref{lan1}) satisfies
\begin{equation}
\left(\frac{\partial}{\partial t}+\nu
k^2\right)G(k;t,t')+\int_{t'}^t ds \,\eta (k;t,s)G(k;s,t')=0
\,\forall \,t>t' \label{lan2}
\end{equation}
and $G(k;t',t')=1$. The expression for $\eta$ given by
\begin{equation}
\eta (k;t,s)=\int d^3{\bf p}L({\bf k},{\bf p})G(p;t,s)Q(|{\bf k
}-{\bf p}|;t,s) \label{eta}
\end{equation}
would make equation (\ref{lan2}) identical to SCF equation
(\ref{scf1}) for the Green's function. To obtain $F(k,t,s)$ we
compare equation (\ref{scf4}) with the equation for $Q(k;t,t')$
obtained from Langevin equation (\ref{lan1}), written as
\begin{equation}
\left(\frac{\partial}{\partial t}+\nu
k^2\right)Q(k;t,t')+\int_{0}^t ds \,\eta
(k;t,s)Q(k;s,t')=\int_{0}^{t'}ds \,G(k;t',s)F(k,t,s). \label{lan3}
\end{equation}
While writing this equation we have made use of equations
(\ref{prop1})-(\ref{prop3}). On comparison and making use of
expression for $\eta$ and generalized fluctuation-dissipation
relation (\ref{scf3}), we obtain
\begin{equation}
F(k,t,s)=\eta (k;t,s)Q(k;s,s) \label{fk}
\end{equation}
which would make (\ref{lan3}) identical to SCF equation
(\ref{scf4}) for $Q(k;t,t')$. Now the equation for $Q(k;t,t)$ as
obtained from the Langevin equation (\ref{lan1}) and using
equations (\ref{prop1})-(\ref{prop3}) can be written as
\begin{equation}
\left(\frac{\partial}{\partial t}+2\nu
k^2\right)Q(k;t,t)+2\int_{0}^t ds \,\eta
(k;t,s)Q(k;t,s)=2\int_{0}^{t}ds
\,G(k;t,s)F(k,t,s)+B(k,t).\label{lan4}
\end{equation}
Comparison of this equation with SCF equation (\ref{scf2}) for
$Q(k;t,t)$, making use of expressions for $\eta (k;t,s)$,
$F(k,t,s)$ given by equations (\ref{eta}) and (\ref{fk})
respectively and using equation (\ref{scf3}) we obtain
\begin{equation}
B(k,t)=2\int d^3{\bf p}L({\bf k},{\bf p})\int_{0}^{t} ds
\,G(k;t,s)Q(|{\bf k }-{\bf p}|;t,s)Q(p;t,s)-2\int_0^t ds\,\eta
(k;t,s)Q(k;t,s) \label{bk}
\end{equation}
which makes equation (\ref{lan4}) identical to equation
(\ref{scf2}). Thus, the Langevin equation (\ref{lan1}) along with
the expression for $\eta$ given by (\ref{eta}) and statistical
properties of the two forcing functions $F(k,t,s)$ and $B(k,t)$
given by (\ref{fk}) and (\ref{bk}) respectively, is the required
model representation for SCF approach of \cite{Herring66}.

\section{Concluding remarks}

A long awaited model representation for self-consistent-field
approach of \cite{Herring66} has been suggested in this paper.
This model is in the form of a Langevin equation having two
statistically independent forcing terms in contrary to one forcing
term present in DIA's Langevin model representation. The proposed
model assures that the closed set of equations of SCF approach
generates statistical properties of the velocity field that is
realizable. It should be noted that the expression for $\eta
(k;t,s)$ is identical to that present in DIA's Langevin model
representation (\cite{Kraichnan70a}). It is worth mentioning here
the reason for different type of model representations for SCF
approach, LET theory and extended Edwards's theory despite the
fact that the generalized fluctuation-dissipation relation is
central to all of them. The SCF approach has been modelled here by
Langevin equation whereas LET and extended Edwards's theories have
an almost-Markovian model representations (\cite{Pandya04},
\cite{Kraichnan71}). This difference is mainly due to an
additional condition for Green's function {\it i.e.}
$G(k;t,t')=G(k;t,s)G(k;s,t')$ which is the property of only LET
and extended Edwards's theories and is satisfied by an
almost-Markovian equation and not satisfied by the Langevin
equation and SCF approach.

\underline{Acknowledgement}

 I acknowledge the financial support provided by
the University of Puerto Rico at Mayaguez, Puerto Rico, USA.

\end{document}